\documentclass[]{aa}  
\usepackage{subcaption}
\usepackage[labelformat=parens,labelsep=quad,skip=3pt]{caption}
\usepackage{graphicx}
\usepackage[varg]{txfonts}
\usepackage{xspace}
\usepackage{color}
\usepackage[]{changes}
\definecolor{DarkGreen}{rgb}{0.0,0.4,0.0}  

\newcommand{\kms}{~$\mathrm{km}\,\mathrm{s}^{-1}$\xspace} 
\newcommand{\skms}[1]{$\sim\,${#1}~km$\,\text{s}^{-1}$} 
\makeatletter
\newcommand*{\rom}[1]{\expandafter\@slowromancap\romannumeral #1@}
\makeatother

\begin{document}

\title{Investigation on the Spatiotemporal Structures of Supra-Arcade Spikes}

\author{Rui Liu\inst{1,2,3}
        \and 
        Yuming Wang\inst{1,2}
        }

\institute{CAS Key Laboratory of Geospace Environment, Department of Geophysics and Planetary Sciences,\\
University of Science and Technology of China, Hefei, 230026, China\\ \email{rliu@ustc.edu.cn}
\and
CAS Center for Excellence in Comparative Planetology,\\
University of Science and Technology of China, Hefei 230026, China
\and
Mengcheng National Geophysical Observatory\\
University of Science and Technology of China, Mengcheng 233500, China
}

\date{Received ???; accepted ???}
\titlerunning{Supra-Arcade Spikes}
\authorrunning{Liu \& Wang}
\abstract
{The vertical current sheet (VCS) trailing coronal mass ejections (CMEs) is the key place where the flare energy release and the CME buildup take place through magnetic reconnection. It is often studied from the edge-on perspective for the morphological similarity with the two-dimensional ``standard'' picture, but its three dimensional structure can only be revealed when the flare arcade is observed side on. The structure and dynamics in the so-called supra-arcade region thus contain important clues to the physical processes in flares and CMEs. } 
{Here we focus on the supra-arcade spikes (SASs), interpreted as the VCS viewed side-on, to study their spatiotemporal structures. By comparing the number of spikes and the in-situ derived magnetic twist in interplanetary CMEs (ICMEs), we intend to check on the inference from the standard picture that each spike represents an active reconnection site, with each episode of reconnection adding approximately one turn of twist to the CME flux rope. } 
{For this investigation we selected four events, in which the flare arcade has a significant north-south orientation and the associated CME is traversed by a near-Earth spacecraft. We studied the SASs using  high-cadence high-resolution 131~{\AA} images from the Atmospheric Imaging Assembly on board the Solar Dynamics Observatory.}
{By identifying each individual spike during the decay phase of the selected eruptive flares, we found that the widths of spikes are log-normal distributed, while the Fourier power spectra of the overall supra-arcade EUV emission, including bright spikes and dark downflows as well as the diffuse background, are power-law distributed, in terms of either spatial frequency $k$ or temporal frequency $\nu$, which reflects the fragmentation of the VCS. We demonstrate that coronal emission-line intensity observations dominated by Kolmogorov turbulence would exhibit a power spectrum of $E(k)\sim k^{-13/3}$ or $E(\nu)\sim \nu^{-7/2}$, which is consistent with our observations. By comparing the number of SASs and the turns of field lines as derived from the ICMEs, we found a consistent axial length of $\sim\,$3.5~AU for three events with a CME speed of \skms{1000} in the inner heliosphere, but a much longer axial length ($\sim\,$8~AU) for the fourth event with an exceptionally fast CME speed of \skms{1500}, suggesting that this ICME is flattened and its `nose' has well passed the Earth when the spacecraft traversed its leg.} 
{}


\maketitle
        
\section{Introduction}

Magnetic reconnection, a fundamental process in plasma, plays a crucial role in solar flares and coronal mass ejections (CMEs). Numerous theoretical models and numerical simulations have demonstrated the formation of a vertical current sheet (VCS) beneath CMEs \citep[see the review by][]{Lin2015}. In the two-dimensional ``standard'' picture \cite[e.g.,][]{Kopp&Pneuman1976,Shibata1995,Lin&Forbes2000,Lin2004}, the VCS forms as a rising flux rope stretches the overlying field, which drives a plasma inflow carrying oppositely directed magnetic field into the VCS region. Magnetic reconnection at the VCS turns the overlying, untwisted field lines that constrain the flux rope into twisted field lines that envelopes the flux rope, which strengthens the upward magnetic pressure force while weakening the downward magnetic tension force, therefore leading to the eruption of the flux rope. While adding layers and layers of hot plasma and twisted magnetic flux to the snowballing flux rope, reconnections at the VCS also produce growing flare loops, whose footpoints in the chromosphere are observed as two flare ribbons separating from each other. In the general three-dimensional geometry, the VCS may form at the cross section of two intersecting quasi-separatrix layers, also termed a hyperbolic flux tube, below the flux rope \citep{Titov2002,Janvier2015,Liu2020}. In addition to the classic two-ribbon flares, complex magnetic topologies may result in diverse ribbon morphologyies, including but not limited to circular ribbon \citep[e.g.,][]{Wang&Liu2012,Chen2020}, X-shaped ribbon \citep[e.g.,][]{Liu2016}, and multi-ribbons \citep[e.g.,][]{Qiu2020}.

In the era before the launch of the Solar Dynamics Observatory \citep[SDO;][]{Pesnell2012}, a major difficulty in studying the VCS lies in the absence of a narrow-band imaging instrument sensitive to hot plasmas above 3 MK. Considerable attention has been given to a coaxial, bright ray feature that appears in coronagraph several hours after some CMEs \citep[e.g.,][]{Webb2003,Lin2005}. Occasionally a bright linear feature in soft X-rays or EUV is observed to extend upward from the top of cusp-shaped flaring loops and line up with the white-light post-CME rays \citep[e.g.,][]{Savage2010,Liu2010,Liu2011}. In the SDO era, the VCS has been recognized not only when it is observed edge on \citep[e.g.,][]{Liu2013,LiuW2013,Zhu2016,Cheng2018,Warren2018,Gou2019}, i.e., when the morphology can be favorably compared with the standard picture; but also when it is observed with a face-on perspective \citep[e.g.,][]{Warren2011, Savage2012reconnection, Savage2012sad,McKenzie2013, Doschek2014, Hanneman+Reeves2014,Innes2014}, i.e., when the post-flare arcade is observed from a line of sight (LOS) perpendicular to its axis. In this case, however, supra-arcade downflows (SADs) catch more attention than any other features. These are tadpole-like dark voids falling at a speed ranging from tens to hundreds of kilometers per second through a fan-shaped, haze-like flare plasma above the flare arcade, which is sometimes described as a fan of ``coronal rays'' \citep{Svestka1998} or ``spike-like rays'' \citep{McKenzie&Hudson1999}, also referred to as the ``supra-arcade fan'' in the literature \citep{Innes2014,Reeves2017}. SADs are generally believed to be a manifestation of reconnection outflows resulting from magnetic reconnection in the VCS. 

\begin{figure*} 
	\centering
	\includegraphics[width=0.95\hsize]{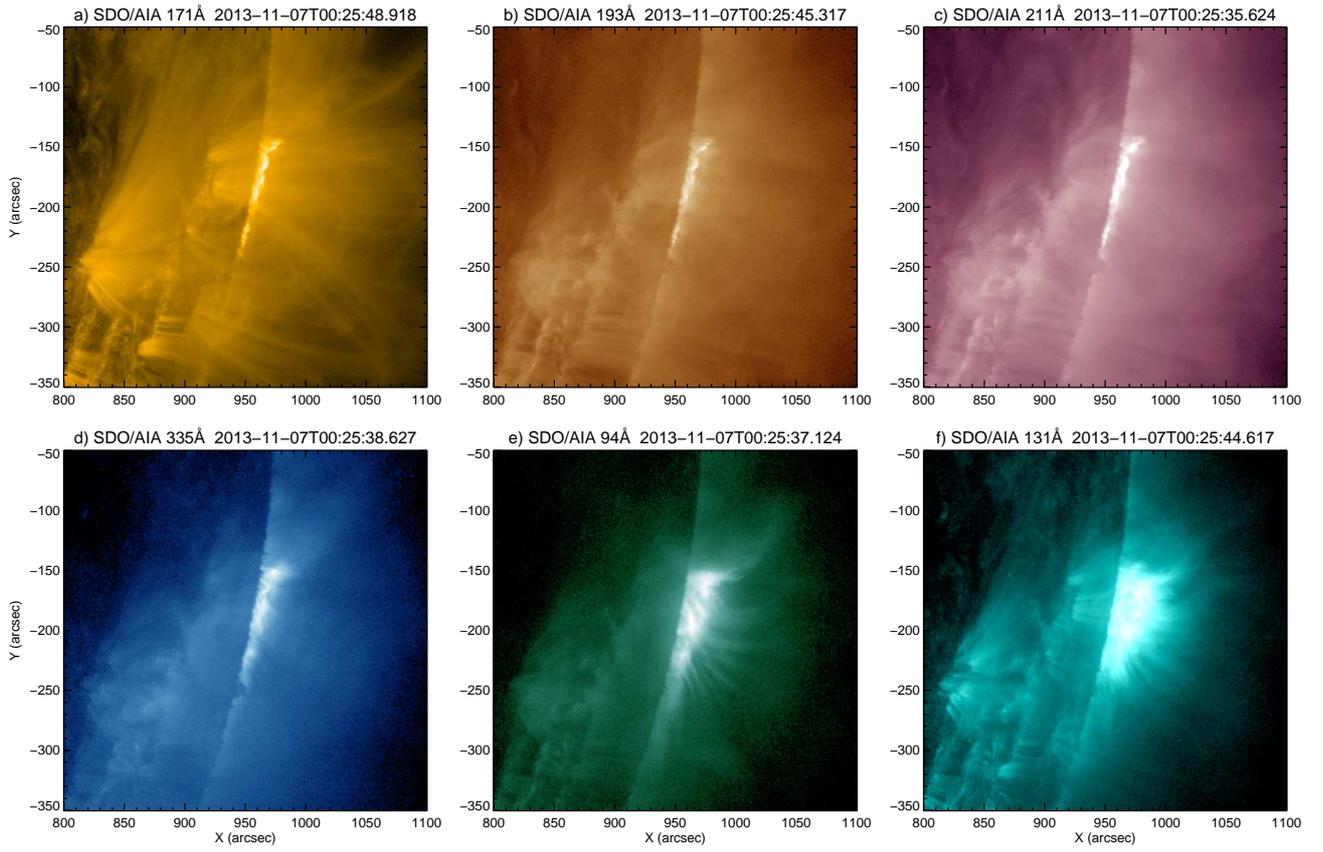}
	\caption{A post-flare arcade observed by different SDO/AIA passbands on 2013 November 7 (Event \#3 in Table~\ref{table:events}). The panels are arranged in an order of increasing characteristic temperatures of the AIA passbands \citep{ODwyer2010}. Note the SASs are only visible in 94 and 131~{\AA}. An animation of AIA 131~{\AA} images is available online. \label{fig:aia}}
\end{figure*}

The VCS must be a three-dimensional structure, which is implied by the post-flare arcade. However, it may not be a continuous ``slab'', extending continuously along the axis of the post-flare arcade, because 1) the arcade consists of numerous discrete post-flare loops \citep[e.g.,][]{Jing2016}; 2) the flare ribbons, which are the footpoints of post-flare loops, often consist of discrete kernels in H$\alpha$ and UV \citep[e.g.,][]{Li&Zhang2015,Lorincik2019}; and 3) the supra-arcade fan is far from uniform but the haze-like plasma is often superimposed by multiple spike-like features (e.g, Figure~\ref{fig:aia}(e \& f)), which are discretely aligned along the top of the arcade and termed supra-arcade spikes (SASs) in this paper. If a post-flare arcade is viewed from an LOS along its axis and there are no prominent coronal features in the foreground or background along the LOS, then the resultant supra-arcade emission must be integrated over the aligned SASs, which gives a linear bright feature extending vertically above the top of the arcade, i.e., a typical VCS \citep[e.g.,][]{Liu2010,Cheng2018}. Both the fan and the SASs are often exclusively observed in the 94 and 131~{\AA} passbands of the Atmospheric Imaging Assembly \citep[AIA;][]{Lemen2012} on board SDO, which are sensitive to plasmas as hot as $\sim\,$6--10~MK (cf. Figure~\ref{fig:aia}). The plasma heating can be naturally attributed to magnetic reconnection in the VCS. 

Here, we carry out an investigation on the spatiotemporal structures of SASs observed during the decay phase of eruptive flares, using high-cadence high-resolution AIA 131~{\AA} images. In the sections that follow, we select four events with typical SAS features (\S~\ref{subsec:select}) and make efforts to identify each spike and estimate its width (\S~\ref{subsec:analysis}). We then study the statistical distribution of spike widths (\S~\ref{subsec:width}) and Fourier power spectra of the supra-arcade EUV emission (\S~\ref{subsec:spectra}), discuss the implication for interplanetary CMEs (ICMEs; \S~\ref{subsec:icme}), and finally summarize the major results (\S~\ref{subsec:summary}). 

\section{Observation \& Analysis}

\subsection{Event Selection \& Overview} \label{subsec:select}
We carried out a survey of flares with magnitude of M-class and above in the SDO era to search for events in which the post-flare arcade is viewed side-on so that the supra-arcade spikes (SASs) are visible. To make certain that a magnetic flux rope is involved in the eruption, we required that the associated CME eventually arrives at Earth and is observed as a magnetic cloud \citep{Burlaga1981,Burlaga1988} by a near-Earth spacecraft. However, such events are relatively rare, because SASs are preferentially visible close to the limb, while CMEs originating from the disk center are more probable to arrive at Earth than those from near the limb. Eventually we found four events as listed in Table~\ref{table:events}; each flare is associated with a halo CME that is observed in situ as an ICME 3--4 days later. 

\begin{table*}
	\caption{List of events} 
	\label{table:events} 
	\centering 
	\begin{tabular}{c c c c c c c p{40mm}} 
	\hline\hline 
	No. & Date & AR & Flare Peak & Flare Class & ICME Date\tablefootmark{a} & ICME $(\theta,\phi)$ \tablefootmark{b} & Reference\\ 
	\hline 
	1 & 2011-Oct-22 & 11314 & 13:09 & M1.3 & 2011-Oct-25 & (36.2, 119.6) & {\cite{Savage2012sad,Reeves2017,Hanneman&Reeves2014,Scott2016,Xue2020}}\\	
	2 & 2013-Apr-11 & 11719 & 07:16 & M6.5 & 2013-Apr-14 & (60.3, 149.9) & {\cite{Samanta2021}} \\ 
	3 & 2013-Nov-07 & 11882 & 00:02 & M1.8 & 2013-Nov-11 & (9.6, 232.4) & \\
	4 & 2014-Apr-02 & 12027 & 14:05 & M6.5 & 2014-Apr-05 & (-0.6, 319.7) & {\cite{Chen2017}} \\
	\hline 
\end{tabular}
\tablefoot{\tablefoottext{a}{Arrival date detected by near-Earth spacecrafts.}\\
	\tablefoottext{b}{Elevation ($\theta$) and azimuthal $\phi$ angles (in degree) of the flux-rope axis in the GSE coordinate system, given by the fitting procedure \citep{Wang2015}.}
}
\end{table*}

The SASs are preferentially observed in 131~{\AA} (Fe~\rom{21} and Fe~\rom{23}), fairly visible in 94~{\AA} (Fe~\rom{18}) due to inferior signal-to-noise ratio and cooler characteristic temperature in this passband, and often invisible in other even ``cooler'' passbands of the Atmospheric Imaging Assembly \citep[AIA;][see Figure~\ref{fig:aia}]{Lemen2012} on board the Solar Dynamics Observatory \citep[SDO;][]{Pesnell2012}. Hence we focuse on the 131~{\AA} passband in this study. The AIA takes full-disk images with a spatial scale of $0''.6$ pixel$^{-1}$ and a cadence of 12 s. In particular, Fe~\rom{21} dominates the 131~{\AA} passband for the flare plasma, with a peak response temperature $\log T = 7.05$, and Fe~\rom{8} dominates for the transition region and quiet corona, with a peak response temperature $\log T = 5.6$ \citep{ODwyer2010}. Previous studies applying the differential emssion measure technique to AIA's six optically thin passbands, i.e., 171, 193, 211, 335, 94, 131~{\AA} (cf. Figure~\ref{fig:aia}) generally demonstrated a significant presence of $\sim\,$10~MK plasma above the post-flare arcade \citep[e.g.,][]{Gou2015}. Table~\ref{table:events} lists publications that study the supra-arcade plasma of the same flares. 

The CMEs are observed by the C2 and C3 coronagraphs of the Large Angle and Spectrometric Coronagraph Experiment \citep[LASCO;][]{Brueckner1995} on board the Solar and Heliospheric Observatory \citep[SOHO;][]{Domingo1995}. The CME speed in the inner heliosphere (within 30 solar radii) is estimated through a linear fitting to the height-time measurements of the CME front and recorded by the SOHO LASCO CME Catalog\footnote{\url{https://cdaw.gsfc.nasa.gov/CME_list/}}. The ICMEs are detected in situ as they traverse the Wind spacecraft near the Earth. We mainly consult the ICME list\footnote{\url{http://space.ustc.edu.cn/dreams/wind_icmes/}} compiled by \citet{Chi2016}. For each ICME, there could be multiple CME candidates within a time window of 36 hrs, i.e., about 3--4.5 days before the ICME arrival at the Earth, roughly corresponding to an average propagation speed of 400--600~\kms between the Sun and Earth. We determine the source of the ICMEby carefully taking into account the CME characteristics such as its location, direction, speed, size, timing, etc. The CME-ICME association for Event~\#2 has been well established by \citet{Vemareddy&Mishra2015}.

\subsection{Data Analysis} \label{subsec:analysis}
\begin{figure*} 
	\centering
	\includegraphics[width=0.95\hsize]{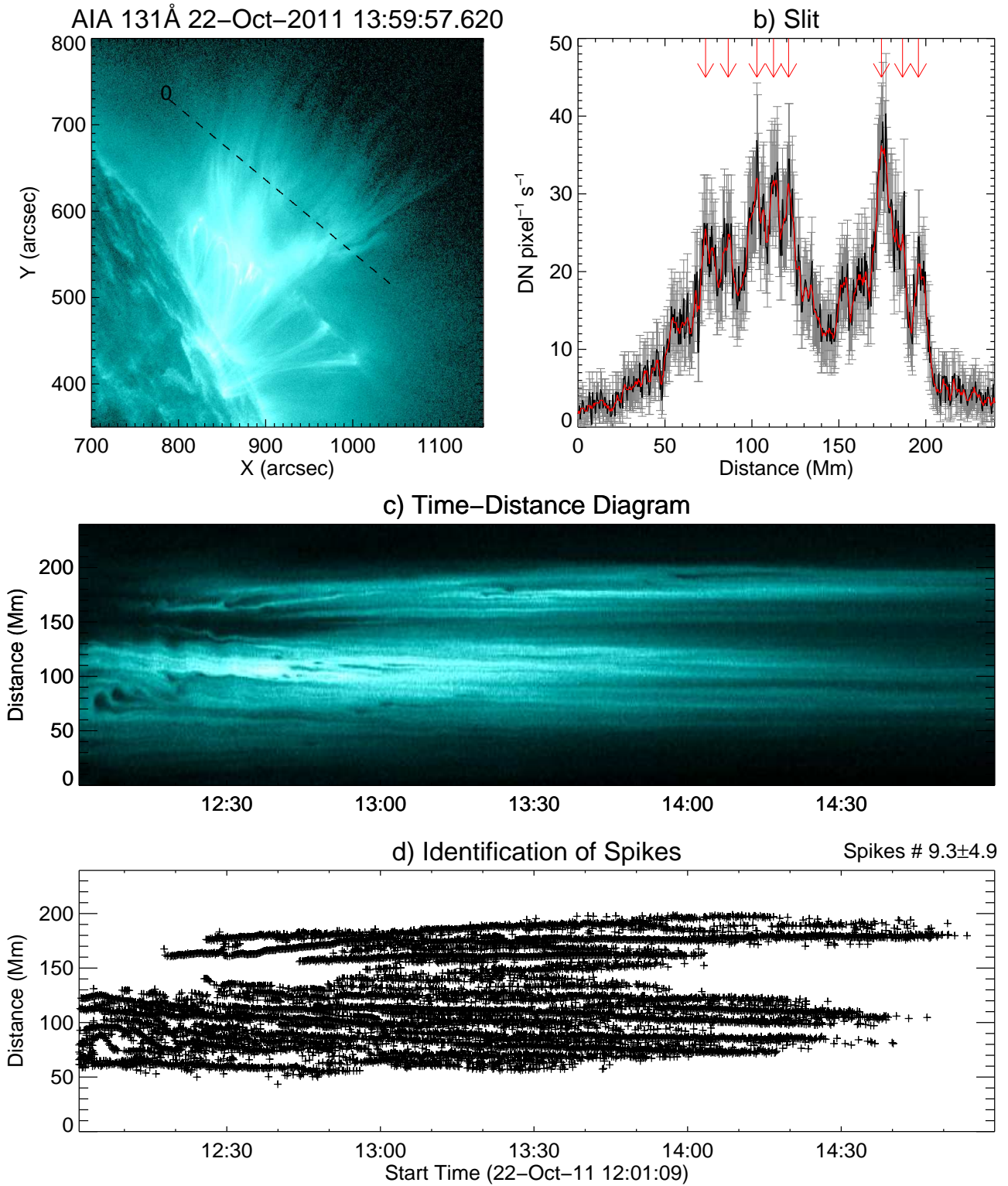}
	\caption{Supra-arcade spikes observed in the 2011 October 22 event. Panel (a) shows a snapshot of the flare in the SDO/AIA 131~{\AA} passband. The brightness in units of $\text{DN}\,\text{pixel}^{-1}\,\text{s}^{-1}$ along a virtual slit (dashed line) is plotted in (b). The starting end of the slit is labeled `0'. In (b) the red curve  shows a five-point running average, the error bars indicate the measurement errors given by \texttt{AIA\_BP\_ESTIMATE\_ERROR}, and the red arrows on the top mark the local maximums identified as the spikes. Panel (c) shows the evolution of the supra-arcade emission seen through the slit in (a). The identified spikes along the slit are marked as crosses in (d), which has the same time and distance axes as  in (c). \label{fig:111022}}
\end{figure*}

\begin{figure*} 
	\centering
	\includegraphics[width=0.95\hsize]{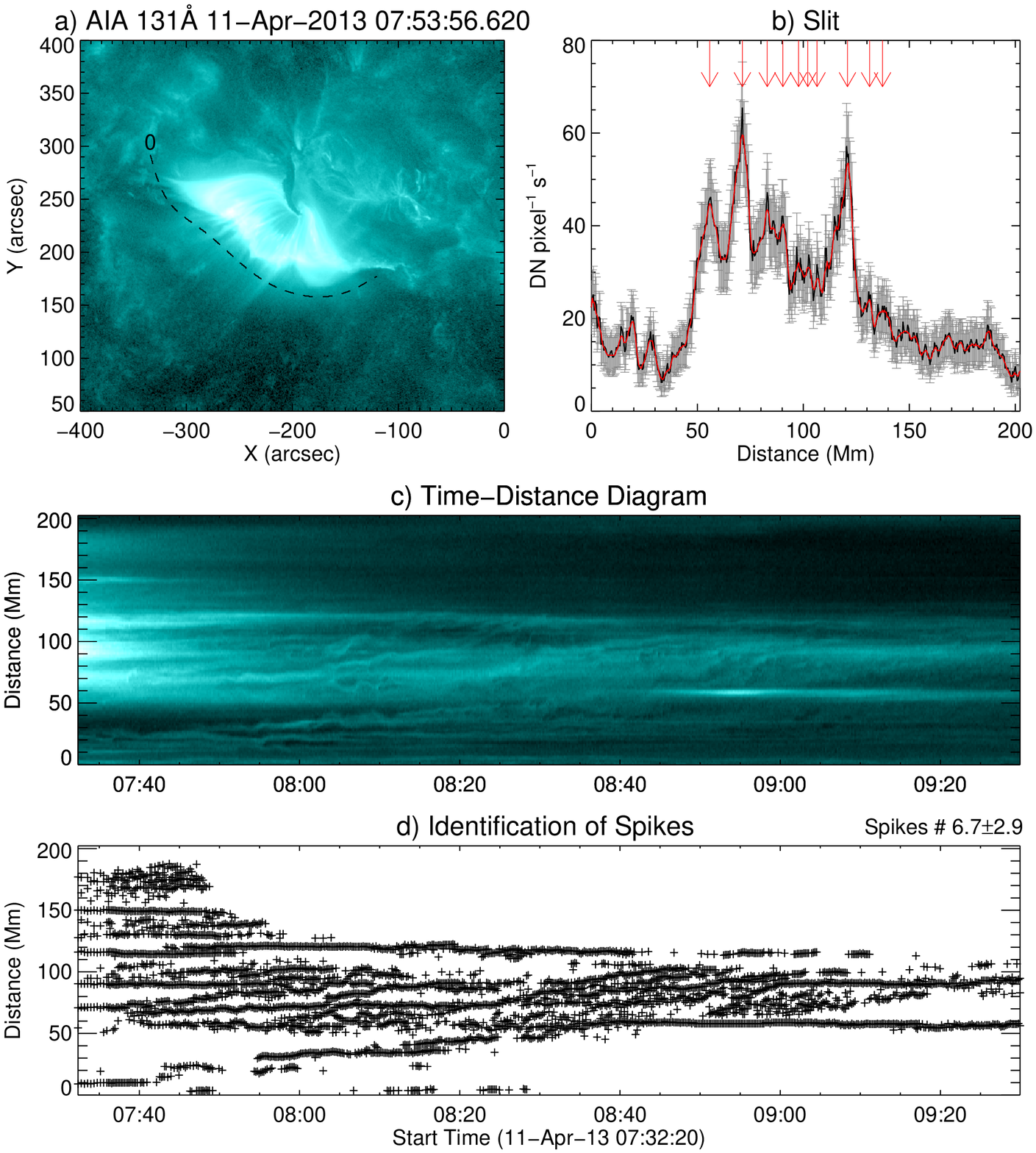}
	\caption{Supra-arcade spikes observed in the 2013 April 11 event. The panels are displayed in the same format as in Figure~\ref{fig:111022}. \label{fig:130411}}
\end{figure*}

Figures~\ref{fig:111022}--\ref{fig:140402} show the spikes detected in a typical timescale of 2 hrs during the flare gradual phase in each of the four events, when the spikes are clearly observed in the AIA 131~{\AA} passband. To detect the spikes, we placed a virtual slit above the post-flare arcade. The slit is either a straight line or a hand-drawn curve to follow the shape of the post-flare arcade (see panel (a) of each figure). Since the brightness along spikes diminishes with height quickly, the slit is placed not far away from the post-flare arcade. The brightness along the slit is shown in panel (b) of each figure.  A visual inspection indicates that each major peak in this plot corresponds to a spike in the image. Note, however, that these major peaks are superposed by rapid wiggles that are comparable to the measurement errors given by \texttt{AIA\_BP\_ESTIMATE\_ERROR} (shown as error bars). To identify the major peaks but discard the wiggles, we first smooth the brightness profile along the slit by a five-point running average (shown as the red curve in (b)), and then locate the local maximums and minimums in a running box of 13 pixels. All local maximums must exceed 20~DN~pixel$^{-1}$ s$^{-1}$, an empirical background set in this study. The distance between two nearest local minimums found at each side of a spike is taken as its width (cf. \S\ref{subsec:width}). The identified major peaks are marked by red arrows on the top of panel (b). 

Further, for the two disk events, \#2 and \#4, AIA 131~{\AA} images are co-registered to the 1st image taken during each selected time interval using the SSW procedure \texttt{DROT\_MAP}. We made the time-distance stack plot (panel (c)) by taking a slice along the virtual slit from each image and then arranging the slices in chronological order. The temporal evolution of spikes is then recorded in this time-distance diagram. One can see that some spikes evolve consistently, some are wavering due to SADs, and others may apparently appear or disappear during the investigated time period. It is interesting that some SADs reveal themselves as dark, tadpole-like features in the time-distance diagram as they propagate through the slit (e.g., Figure~\ref{fig:111022}c). It can be seen that SADs may flow not only along intra-spike ``channels'' but also into the spikes, the latter of which may result in the apparent splitting of a spike (some clear examples can be seen in Figure~\ref{fig:130411}c). Such interactions between bright spikes and dark down-flowing voids are suggested to be evidence for Rayleigh–Taylor type instabilities in the exhaust of reconnection jets \citep{Guo2014,Innes2014}. 

The identified spikes along the slit are plotted as crosses in Panel (d), which has the same time and distance axes as the time-distance diagram in Panel (c). We find our identification algorithm to be satisfying though a visual comparison between (c) and (d). The time-averaged number of spikes and the standard deviation are annotated on the top right of panel (d).   

\section{Discussion}

\subsection{Distribution of spike widths} \label{subsec:width}
With the width of each spike being estimated (\S\ref{subsec:analysis}), we plot the distribution of spike{\tiny 
} widths during the gradual phase of each flare (Figure~\ref{fig:sw}). The distribution of all four flares investigated is rather similar, with a peak at around 5--6~Mm, an average of 16--21 Mm, a standard deviation of 7--11 Mm, and a median of 15--19 Mm. The peak of 5--6~Mm is well consistent with the current-sheet width detected in the lower corona in EUV passbands as reported in the literature \citep[e.g.,][]{Liu2010,Zhu2016}. Each distribution also has a similar lower cutoff of about 3 Mm and an upper limit of about 20~Mm. The lower cutoff can be attributed to the running box of 13 pixels specified in our algorithm, which effectively limits the spike width to be above 6 pixels, or 2.6 Mm. This is smaller than the lower limit of the current-sheet width, 3~Mm, deduced by \citet{Guo2013} from the nonlinear scaling law of the plasmoid instability. Note \citet{Lin2007,Lin2009} gave a larger lower limit (50~Mm) based on the linear scaling law. 

Apparently, spike widths are not normally but log-normally distributed. In Figure~\ref{fig:sw},  employing a nonlinear least squares fitting procedure \texttt{MPFIT}\footnote{\url{http://purl.com/net/mpfit}} \citep{Markwardt2009} and assuming a 2-pixel measurement error,  we fit each distribution with the following function,
\begin{equation} \label{eq:lognorm}
y(x)=\frac{C}{x\sigma\sqrt{2\pi}}\exp\left[-\frac{(\ln x -\mu)^2}{2\sigma^2}\right],
\end{equation} where $C$ is an arbitrary constant. The other two parameters $(\mu,\sigma)$ are annotated in the middle right of each panel. It is noteworthy that we obtain similar $\mu$ and $\sigma$ in all four events.

\subsection{Spatiotemporal structures of supra-arcade EUV emission} \label{subsec:spectra}
\begin{figure*} 
	\centering
	\includegraphics[width=0.95\hsize]{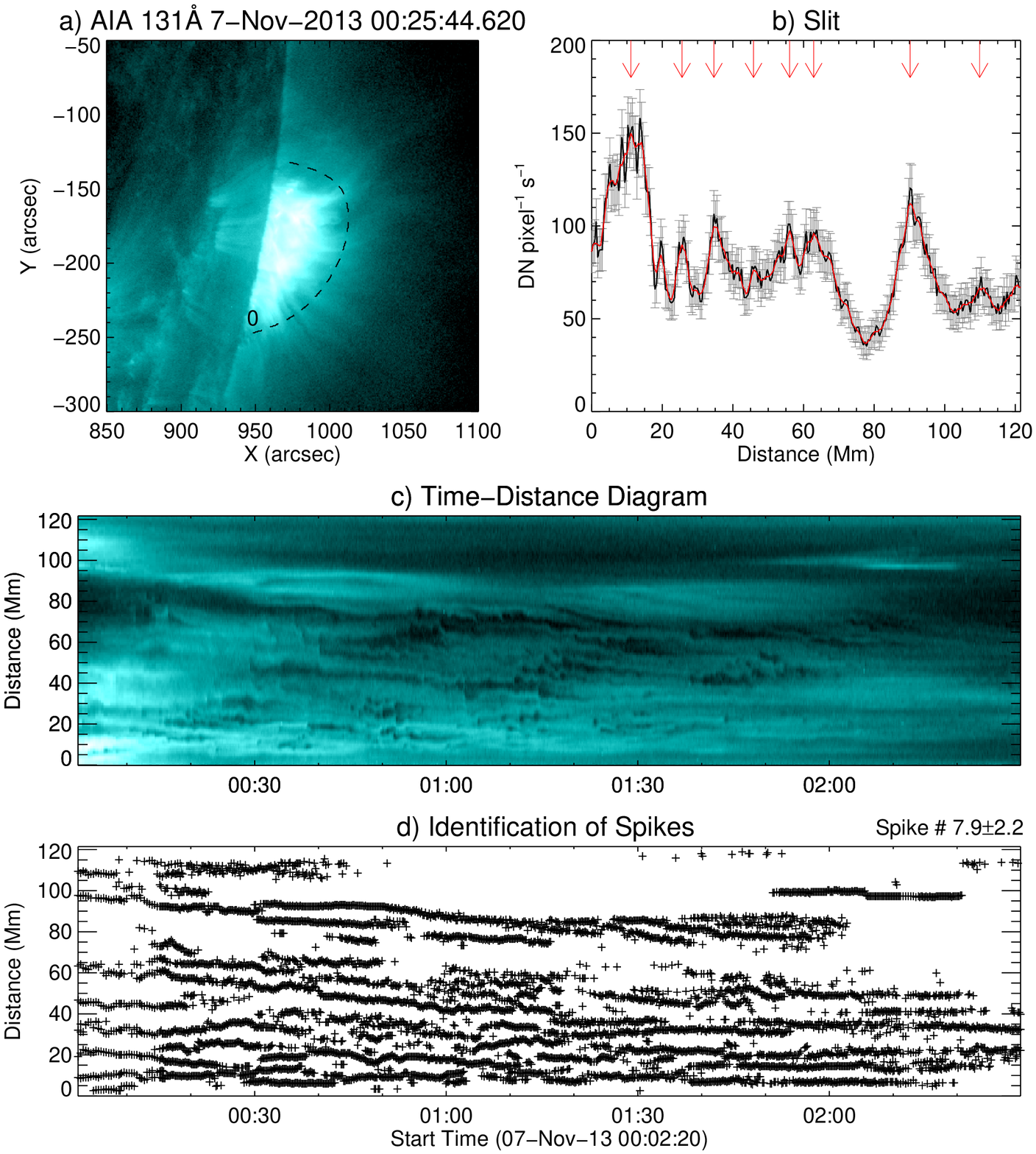}
	\caption{Supra-arcade spikes observed in the 2013 November 7 event. The panels are displayed in the same format as in Figure~\ref{fig:111022}. \label{fig:131106}}
\end{figure*}

\begin{figure*} 
	\centering
	\includegraphics[width=0.95\hsize]{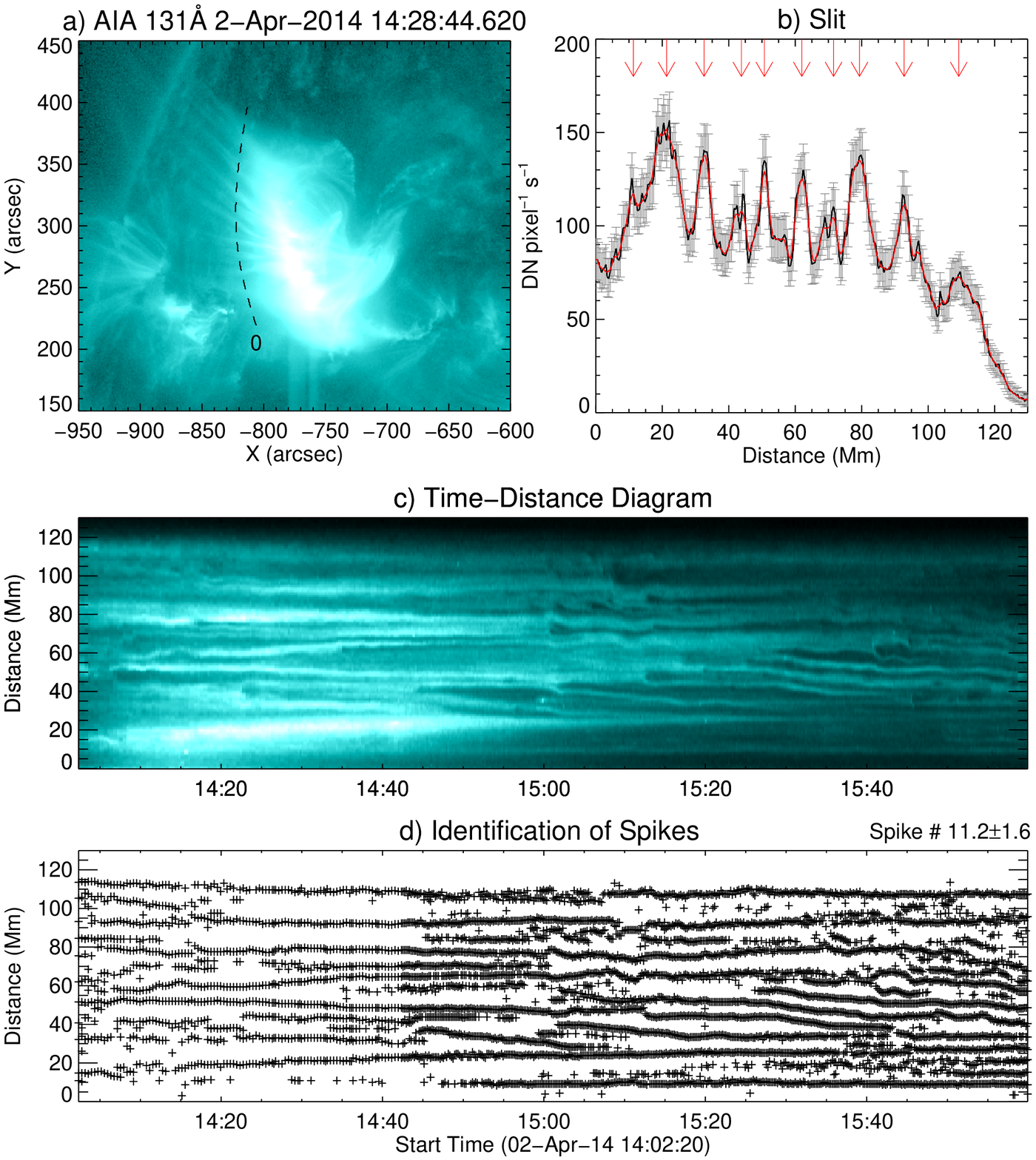}
	\caption{Supra-arcade spikes observed in the 2014 April 2 event. The panels are displayed in the same format as in Figure~\ref{fig:111022}. \label{fig:140402}}
\end{figure*}

\begin{figure} 
	\centering
	\includegraphics[width=\hsize]{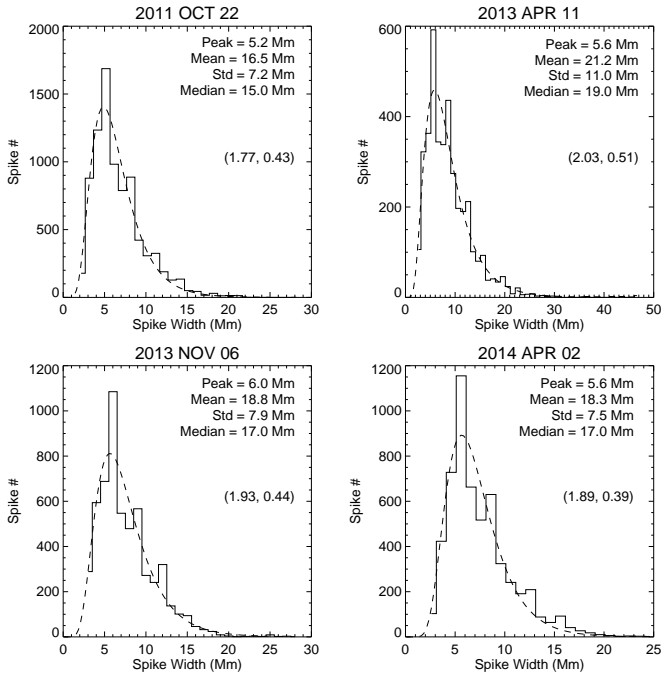}
	\caption{Distribution of the widths of supra-arcade spikes. Each panel shows the histogram for one of the four events under investigation. The statistics of the distribution are indicated in the upper right, and the fitting parameters $(\mu,\sigma)$ of the log-normal function (Eq.~\ref{eq:lognorm}) in the middle right. The  fitting function is plotted as a dashed curve. \label{fig:sw}}
\end{figure}

\begin{figure} 
	\centering
	\includegraphics[width=\hsize]{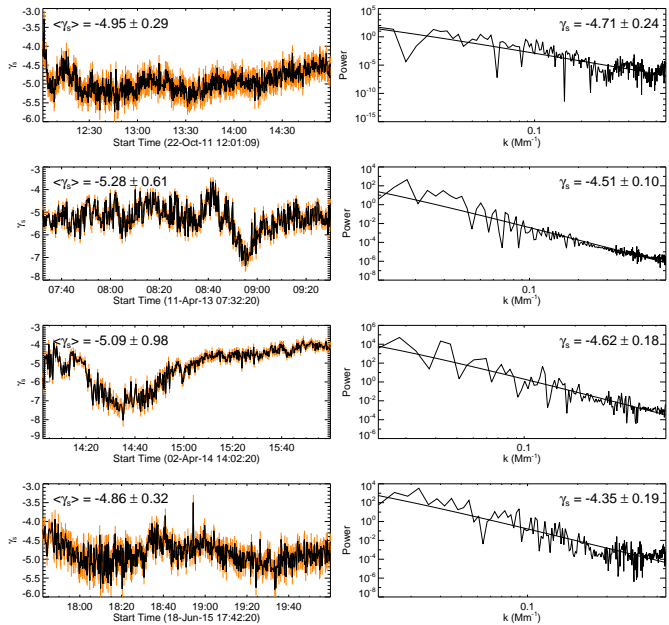}
	\caption{Structure of supra-arcade EUV emission in terms of spatial frequency. The power spectrum of any vertical slice in the time-distance diagram (panel (c) of Figures~\ref{fig:111022}--\ref{fig:140402}) is fitted with a power-law function. The temporal variation of the power-law index $\gamma_s$ is shown in the left panels. The orange bars indicate the 1-$\sigma$ uncertainty estimates in the fittings. The power spectrum of the time-distance diagram collapsed along the time axis is shown in the right panels.  \label{fig:power_s}}
\end{figure}

We apply the Fourier transform to each vertical slice in the time-distance diagram in Panel (c) of Figures~\ref{fig:111022}--\ref{fig:140402} to explore the structure of the supra-arcade emission in terms of spatial frequency $k$ (Mm$^{-1}$). We fit the Fourier power spectrum resulting from each vertical slice with a power-law function and show the temporal variation of the power-law index $\gamma_s$ in the left column of Figure~\ref{fig:power_s}. Collapsing the time-distance diagram along the time axis gives the time-integrated EUV emission profile along the slit. Applying the Fourier transform to this profile, we obtain power spectra exhibiting clear power-law behaviors (right column of Figure~\ref{fig:power_s}), whose fitted indices are in the range of $\sim\,$[4, 5], similar to the range of $\gamma_s$ in the left column. 

\begin{figure} 
	\centering
	\includegraphics[width=\hsize]{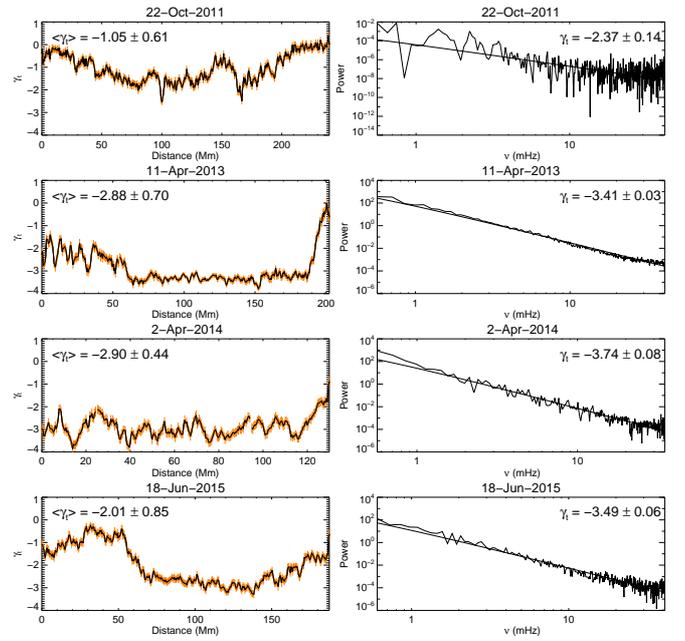}
	\caption{Structure of supra-arcade EUV emission in terms of temporal frequency. The power spectrum of any horizontal slice in the time-distance diagram (panel (c) of Figures~\ref{fig:111022}--\ref{fig:140402}) is fitted with a power-law function. The spatial variation of the power-law index $\gamma_t$ along each slit (panel (a) in Figures~\ref{fig:111022}--\ref{fig:140402}) is shown in the left panels. The orange bars indicate the 1-$\sigma$ uncertainty estimates in the fittings. The power spectrum of the time-distance diagram collapsed along the distance axis is shown in the right panels.  \label{fig:power_t}}
\end{figure}

Similarly, we apply the Fourier transform to each horizontal slice in the time-distance diagram in Panel (c) of Figures~\ref{fig:111022}--\ref{fig:140402} to explore the structure of the supra-arcade emission in terms of temporal frequency $\nu$ (mHz). Fitting the Fourier power spectrum from each horizontal slice by a power-law function, we show the variation of the resultant power-law index $\gamma_t$ along the virtual slit in the left column of Figure~\ref{fig:power_t}. Collapsing the time-distance diagram along the distance axis, we effectively integrate the EUV emission along the slit, which emulates the situation when the VCS is observed edge-on in the optically thin corona. This results in a time series of EUV emission sampled from a point on the vertical current sheet. The power spectra of spatially integrated EUV emission also exhibit clear power-law shapes (right column of Figure~\ref{fig:power_t}), whose indices are in the range of $\sim\,$[2.5, 4], similar to the range of $\gamma_t$ in the left column. 

\citet{Cheng2018} construct time-distance diagrams from a slit along a VCS observed edge on. Sampling two horizontal slices above the flare arcade in a time-distance diagram, they obtained the corresponding Fourier power spectra with a power-law indices $\alpha$ of 1.8 and 1.2, respectively. The difference in power-law indices between $\gamma_t$ in our case and $\alpha$ in \citet{Cheng2018} can probably be attributed to the following differences: 1) the pow spectra in \citet{Cheng2018} appear very noisy (their Figure 13), lacking a strict power-law shape, compared with the nice power-law behavior demonstrated in Figures~\ref{fig:power_s} and \ref{fig:power_t}; and 2) the line-of-sight integration in their case includes the coronal foreground and background, whereas in our case the spatial integration is limited to coronal emission right above the post-flare arcade, therefore less interferences from irrelevant coronal structures.

Below we demonstrate heuristically the consistency of the observed power-law behavior with Kolmogorov turbulence in the supra-arcade region. Any turbulent physical quantity can be written as
\begin{equation}
u(x,t)=u_0(x,t)+\delta u(x,t), \label{eq:turbulent_quantity}
\end{equation}
where $u_0(x,t)$ varies smoothly at large spatiotemporal scales, while $\delta u (x, t)$ fluctuates randomly at small scales, i.e., there is a large separation of both space and time scales between the ``mean'' and fluctuating components. The two components hence respond distinctively to a long wavelength average, i.e.,
\begin{equation}
\langle u_0(x,t)\rangle=u_0(x,t) \quad \text{and} \quad \langle\delta u(x,t)\rangle=0.
\end{equation}

The Fourier transform of the observed intensity $I=I_0+\delta I $ gives the power spectra,
\begin{equation}
E(k)\sim k^{-\gamma_s} \quad\text{and}\quad E(\nu)\sim \nu^{-\gamma_t},
\end{equation} in terms of spatial frequency $k$ and temporal frequency $\nu$, respectively, as shown in Figures~\ref{fig:power_s} and \ref{fig:power_t}. Here for clarity, the spectral index $\gamma_s$ and $\gamma_t$ are set as positive. $I_0$ is transformed to the low-frequency end of the power spectra, and $\delta I$ to the higher-frequency portion with power-law behavior, i.e., $\delta I^2\sim E(k)k \sim k^{-\gamma_s+1} \sim \mathnormal{l}^{\,\gamma_s-1}$, which gives
\begin{equation} \label{eq:deltaI_l}
\delta I \sim l^{\,(\gamma_s-1)/2} \sim l^{\,[3/2,\ 2]},
\end{equation}
because $\gamma_s$ is found to be in the range of [4, 5]. Similarly, since $\delta I^2\sim E(\nu)\nu$, we have
\begin{equation} \label{eq:deltaI_f}
\delta I \sim \nu^{-(\gamma_t-1)/2} \sim \nu^{-[3/4,\ 3/2]},
\end{equation} with $\gamma_t$ in the range of [2.5, 4].

On the other hand, the intensity of coronal emission lines $I\sim \int n^2 ds$ \citep[][\S2.8]{Aschwanden2005}, where $s$ is the distance along the LOS. Expanding $n$ according to Eq.~\ref{eq:turbulent_quantity}, we have
\begin{equation}
I\sim\int (n_0^2+2n_0\delta n+\delta n^2)\,ds.
\end{equation} The LOS integration \emph{a priori} takes a long wavelength average, which gives
\begin{equation}
I_0 \sim \int n_0^2\,ds \quad \text{and} \quad \delta I \sim \int \delta n^2\,ds.  
\end{equation} As a comparison, we anticipate that the turbulent component would be smoothened out in white-light observations of the K-corona, in which the observed intensity is dominated by Thomson scattering, i.e., $I\sim\int n ds$.

The density perturbation $\delta n$ is linked to the velocity fluctuation $\delta \mathbf{u}$ by the equation of continuation
\begin{equation}
\frac{\partial\delta n}{\partial t} \simeq  -\nabla\cdot(n_0\delta \mathbf{u}),
\end{equation} assuming that $|\delta n| \ll n_0$ but $|\partial n_0/\partial t|\ll |\partial\delta n/\partial t|$ and $\mathbf{u}_0\simeq 0$. Thus,
\begin{equation}
\delta n \sim -\frac{n_0}{L_0/t_0}\delta u \sim -n_0\frac{\delta u}{V_A}
\end{equation} in orders of magnitude, where $t_0$ and $L_0$ are characteristic time and length scale of the system, respectively. The assumption that the Alfv\'{e}n speed $V_A$ is on the same order of magnitude as $L_0/t_0$ implies that $|\delta u|\ll V_A$. This is consistent with spectroscopic observations of hot emission lines in the VCS, which give nonthermal velocities of no larger than 200~\kms \citep[e.g.,][]{Li2018}. We further argue that the LOS integration from the side-on perspective of the VCS make relevant the size $l$ of large eddies in the flow velocity field $\mathbf{u}\simeq\delta \mathbf{u}$, with the thickness of the VCS being comparable to those of spikes; i.e., $\delta I\sim \delta u^2\,l$, with $l\lesssim L_0$. $l$ can be deemed as the order of magnitude of distances over which $\delta u$ changes appreciably. 

If the observed intensity perturbations $\delta I$ result from a fully developed turbulence following the famous universal law of Kolmogorov and Obukhov, i.e., $\delta u\sim\mathnormal{l}^{1/3}$ \citep[][\S33]{Landau}, then we have \begin{equation}
\delta I \sim l^{5/3}.
\end{equation} The index 5/3 falls nicely in the compact range of [3/2, 2] as given independently by Eq.~\ref{eq:deltaI_l}. With $\nu\sim (l\,\delta u)^{-1} \sim l^{-4/3}$, we have
\begin{equation}
\delta I \sim \nu^{-5/4},
\end{equation} in which the index also falls in the compact range of [3/4, 3/2] as given by Eq.~\ref{eq:deltaI_f}. 

Thus, when the fluctuations in coronal emission-line intensity observations are dominated by turbulence, the corresponding power spectra are expected to exhibit the power-law behavior as follows, 
\begin{equation}
E(k)\sim k^{-13/3} \quad\text{and}\quad E(\nu)\sim \nu^{-7/2}. 
\end{equation} As a further evidence of turbulence in the VCS region, one can see from the left column of Figure~\ref{fig:power_t} that the power-law index $\gamma_t$ is generally closer to the value of 7/2 in the middle of the slit, that is, within the VCS region, than at the two ends of the slit, that is, outside of the VCS.

In contrast, the LOS integration from the edge-on perspective of the VCS gives $\delta I\sim \delta u^2\,L$, with the `depth' of the VCS being comparable to the axial length of the post-flare arcade, i.e., $L\gg L_0$. This leads to power spectra in $k^{-7/3}$ or $\nu^{-2}$, which are only slightly steeper than those obtained by \citet{Cheng2018}. 

\subsection{Implication for ICMEs} \label{subsec:icme}
Each SAS can be considered as a reconnection site on the 3D VCS in the wake of the CME. According to the standard model, each episode of magnetic reconnection of the overlying field at the VCS adds a twist of roughly 1 turn to the burgeoning CME flux rope. If magnetic reconnections at each SAS proceeds continually, i.e., there is no significant time interval between any two episodes of magnetic reconnection on the same SAS, then at any time one may assume that the number of ongoing reconnection episodes roughly correspond to the number of SASs. Since the number of SASs is stable during the flare decay phase, we expect that the flux rope's outer shells are more or less uniformly twisted \cite[e.g.,][]{Wang2017}. When the CME arrives at the Earth, one would expect near-Earth spacecrafts to detect a magnetic cloud possessing magnetic twist of similar number of turns in its outer shells as the number of SASs.  

\begin{figure*} 
	\centering
	\includegraphics[width=0.95\hsize]{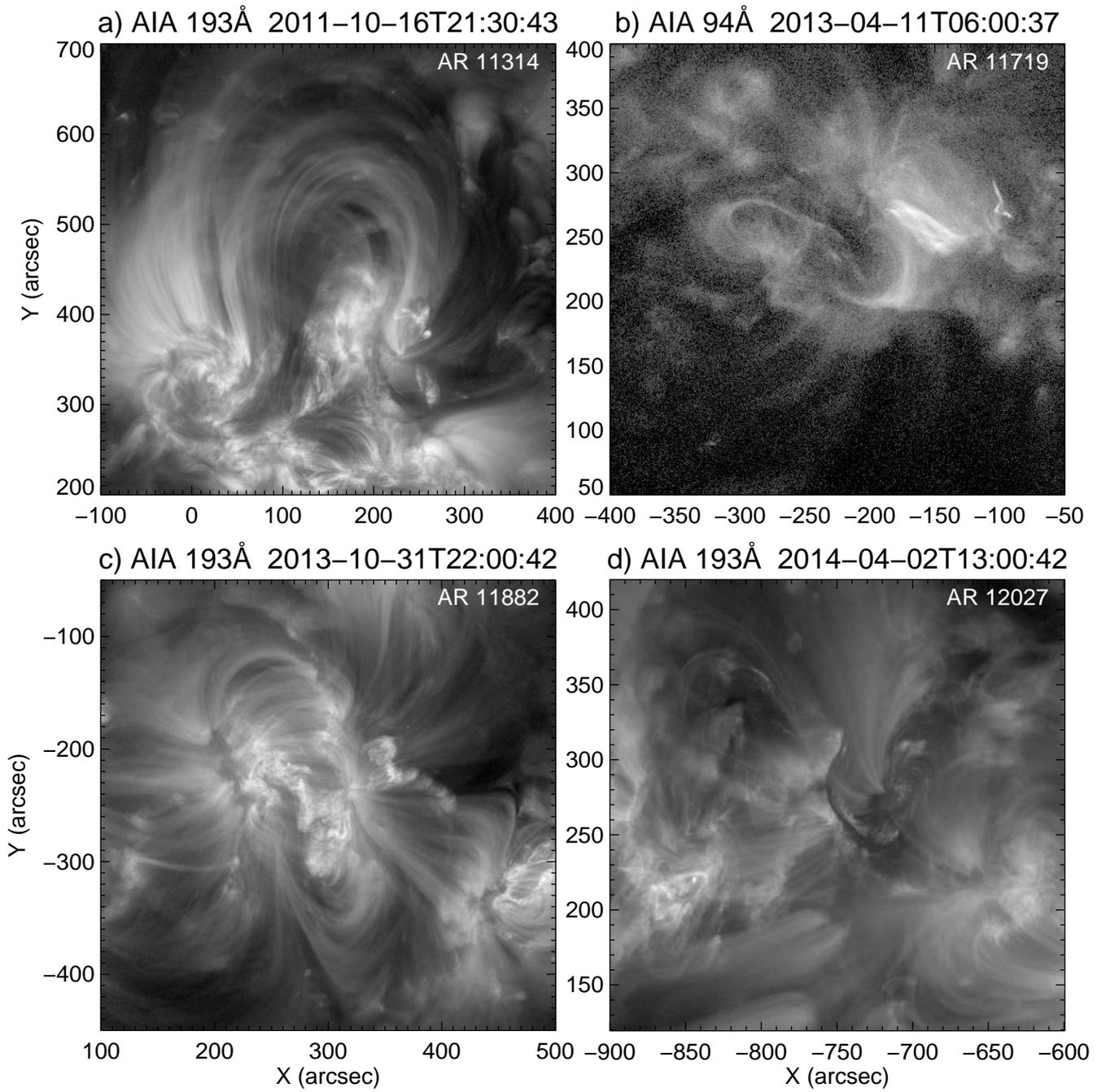}
	\caption{The four active regions under investigation as observed by SDO/AIA. \label{fig:ar}}
\end{figure*}

We fit the four ICMEs by a velocity-modified uniform-twist force-free flux rope model \citep{Wang2016}. The in-situ observations and corresponding fitting results are shown in the Appendix Figures~\ref{fig:mc111022}--\ref{fig:mc140402}. The model yields the twist density $\tau=rB_\phi/B_z$ in local cylindrical coordinates $(r, \phi, z)$ with the $z$-axis directed along the flux-rope axis. $\tau$ is given in units of the number of field-line turns per unit AU along the flux-rope axis. The helicity (twist) signs are negative for all four events, which, except Event \#3, originate from the northern hemisphere, consistent with the cycle-independent dominance of negative helicity in the same hemisphere \citep[][and references therein]{Zhou2020}. Event \#3 originate from a decayed active region (AR 11882) in the southern hemisphere close to the equator. When it is located near the disk center, the active region exhibits a weak revserse-S geometry manifested by coronal loops in EUV (Figure~\ref{fig:ar}(c)), associated with a filament aligned along the polarity inversion line with a similar geometry, signaling the presence of negative helicity. In contrast, AR 11314, the source region of Event \#1 exhibits a forward-S geometry in the east and a reverse-S geomtry in the west (Figure~\ref{fig:ar}(a)), suggesting that the CME probably originates from the west side of the active region. The other two source regions, ARs 11719 (Figure~\ref{fig:ar}(b)) and 12027 (Figure~\ref{fig:ar}(d)), are clearly reverse-S shaped, with the latter being outlined by a sigmoidal filament. Additionally, different aspects of the source region of Event \#2, AR 11719, has been studied in detail by several authors \citep{Vemareddy&Mishra2015,Vemareddy2016,Guo2019}.

The ICME's axial length is unknown but assumed to be lie in the range of 2--$\pi$ AU. At the lower end, the flux rope is so stretched that it is ``folded'' in half between the Sun and Earth; but at the higher end, the rope is circular \citep{Wang2015}. Alternatively, \citet{Hu2015} compared the lengths of ICME field lines that are derived from energetic electron bursts observed at 1 AU and the magnetic twist derived from the Grad-Shafranov reconstruction method. They concluded that the effective axial lengths of the selected ICME events range between 2 to 4 AU. Unfortunately none of our selected events is associated with in-situ energetic electron bursts. However, the comparison between the number of SASs and the magnetic twist of ICMEs as below may shed new light on their axial lengths. 

\begin{figure*} 
	\centering
	\includegraphics[width=\hsize]{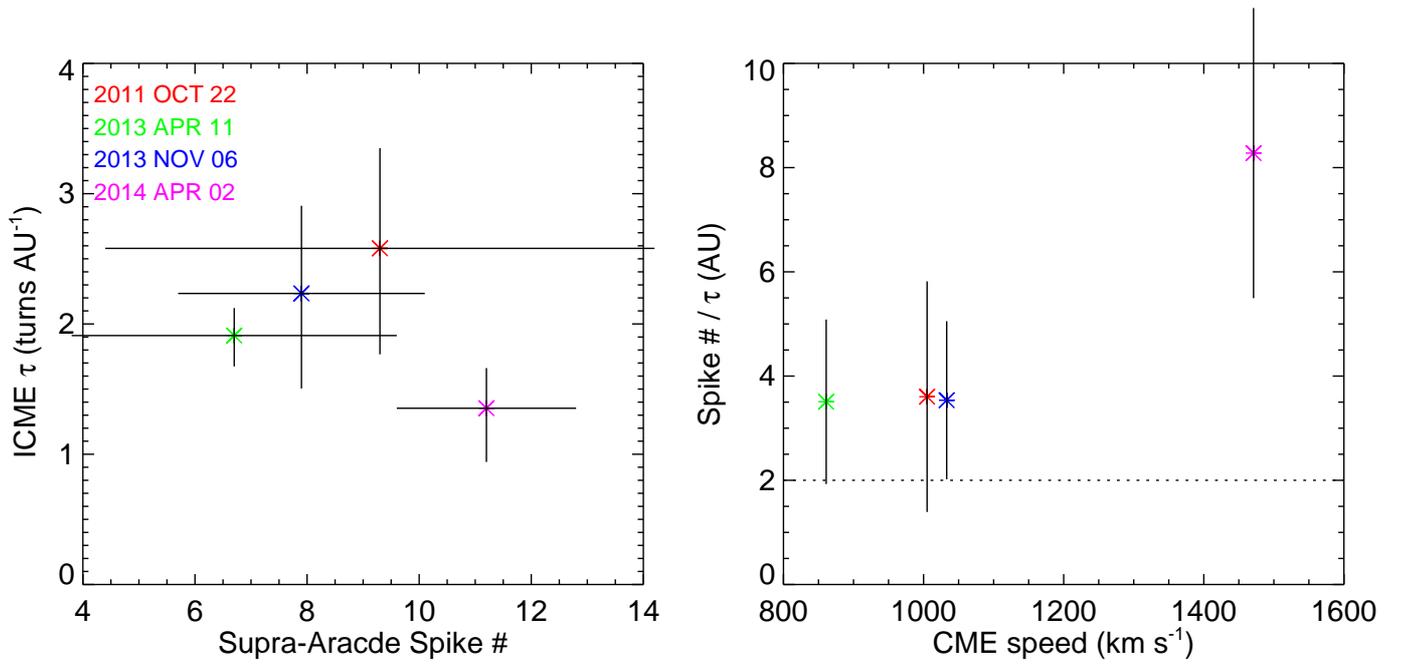}
	\caption{Implication of the number of supra-arcade spikes for ICMEs. Left panel shows the twist density $\tau$ of ICMEs versus the number of spikes. The error bars for the former are given by the fitting of ICMEs with a uniformly twisted flux-rope model (see the text for details), and those for the latter is the standard deviation of the average number of spikes detected during the selected time intervals during the flare decay phase of each event. Right panel shows the ratio of the number of spikes over $\tau$ in relation to the CME speed estimated in the inner heliosphere. The error bars are given by the propagation of uncertainties. 
    \label{fig:comp}}
\end{figure*}

In the first three events, we found that $\tau$ linearly increases with the increasing number of SASs (left panel of Figure~\ref{fig:comp}). Further, the ratio between the number of SASs over $\tau$ gives a similar length of the flux-rope axis, i.e., $\sim\,$3.5~AU (middle panel of Figure~\ref{fig:comp}), which is slightly above $\pi$~AU, but still falls in the range suggested by \citet{Hu2015}. However, the 4th event seem to be an outlier: it is the only event originating from close to the east limb (about $60^\circ$ east), but the resultant ICME is still `skimmed' by the near-Earth spacecraft for over 30 hrs (Appendix Figure~\ref{fig:mc140402}), indicating its large size; the ICME has a moderate $\tau$ but the number of SASs is the largest among the four events, which is translated to an axial length of 8.3~AU. These may be caused by two different but interrelated effects, both resulting from its exceptionally fast speed (\skms{1500} projected on the plane of sky) as measured in the inner heliosphere by SOHO/LASCO. 

First, the ICME experiences a much stronger resistance than the other three events when it `plows' through the solar wind, since in principle the aerodynamic drag force applied on the ICME is proportional to the square of the difference between the ICME speed and the solar wind speed \citep{Cargill2004}. Hence the drag force acting on the fourth ICME is roughly 3.5 times more than that on the other three, given the solar wind speed in the range of 400--450~\kms. The resistance restricts the ICME's radial much more than its lateral expansion \cite[e.g.,][]{Manchester2004}. However, this would lead to an unrealistically highly flattened shape (right panel of Figure~\ref{fig:comp}) if the spacecraft crossed the `nose' of the flux rope. We then suggest that the spacecraft actually traverses the leg of the ICME, which is evidenced by its axis being oriented almost along the Sun-Earth line (Table~\ref{table:events}). As a comparison, the flux-rope axes in the other three events make large angles with respect to the Sun-Earth line. Hence at the time of the spacecraft passing, the `nose' of the fourth ICME has well passed the Earth, resulting in an axial length much longer than $\pi$ AU. However, if the axis has a circular shape, the nose of the ICME would be over 2.5~AU from the Sun. This is translated to an average propagation speed of \skms{1300}, which is also unrealistic. In reality, we may be witnessing a combination of both the flattening and the leg-passing effect.

\citet{Wang2016} converted the field-line length $L_\text{mfl}$ estimated by the electron-probe method in seven magnetic clouds from \citet{Kahler2011} to a twist density $\tau_e$ based on the uniformly twisted Gold-Hoyle flux rope model \citep{Gold&Hoyle1960}, i.e., $L_\text{mfl}=L_a\sqrt{(1+(2\pi\tau_e r)^2}$, assuming that the flux-rope axial length $L_a$ is in the range of 2--$\pi$ AU. A good correlation is found between  $\tau_e$ and $\tau$ except for an outlier with an exceptionally large $\tau_e$ exceeding 10 turns~AU$^{-1}$. This could result from an underestimated $L_a$, in light of our Event \#4, whose estimated axial length greatly exceeds $\pi$ AU. 

\section{Conclusion} \label{subsec:summary}
To summarize, we investigate the SASs observed during the decay phase of four selected eruptive flares, using high-cadence high-resolution AIA 131~{\AA} images. By identifying each individual spike, we find that the spike widths are log-normally distributed. The distribution has a lower cutoff at 2.6~Mm due to our identification algorithm. On the other hand, the Fourier power spectra of the overall supra-arcade EUV emission, including spikes, downflows, and haze, are power-law distributed, in terms of both spatial and temporal frequencies up to $\sim\,$0.7~Mm$^{-1}$ and $\sim\,$40~mHz, respectively. Further, assuming that each spike represents an active reconnection site on the three-dimensional VCS, we compare the number of SASs with the turns of field lines as derived from the ICMEs. The comparison gives a consistent axial length of $\sim\,$3.5~AU for three events with a CME speed of \skms{1000} in the inner heliosphere, but a much longer axial length ($\sim\,$8~AU) for the fourth event with a CME speed of \skms{1500}, which may be a consequence of a combination of the flattening and the leg-crossing effect. 

The power-law behavior of the supra-arcade power spectra demonstrates the absence of a dominant spatio-temporal scale in the supra-arcade region, and furthermore, the power-law indices are consistent with Kolmogrov turbulence developed in the VCS, probably resulting from, e.g., the plasmoid instability \citep{Loureiro&Udensky2016}. It is clear from the side-on perspective that not only the bright spikes but also the diffuse background and dark voids collectively contribute to the power-law behavior. The widths of spikes are log-normally distributed, which might be a signature of the fragmentation of the VCS, since the log-normal distribution is known to be able to well describe the size distribution in a wide variety of fragmentation phenomena \citep[e.g.,][]{Kolmogorov1941,Cheng&Redner1988,Brown1989}.

In addition, to verify the axial length as inferred from the ratio between the number of SASs and the ICME's twist density, we need look for further constraints, such as those from interplanetary energetic electrons or multi-spacecraft observations, in future investigations.

\bibliographystyle{aa}
\bibliography{cs}
\begin{acknowledgements}
This work was supported by the Strategic Priority Program of the Chinese Academy of Sciences (XDB41030100) and the National Natural Science Foundation of China (NSFC; 41774150 and 11925302). 
\end{acknowledgements}

\begin{appendix}
	\section{Supplementary figures}

\begin{figure*} 
	\centering
	\includegraphics[width=0.8\hsize]{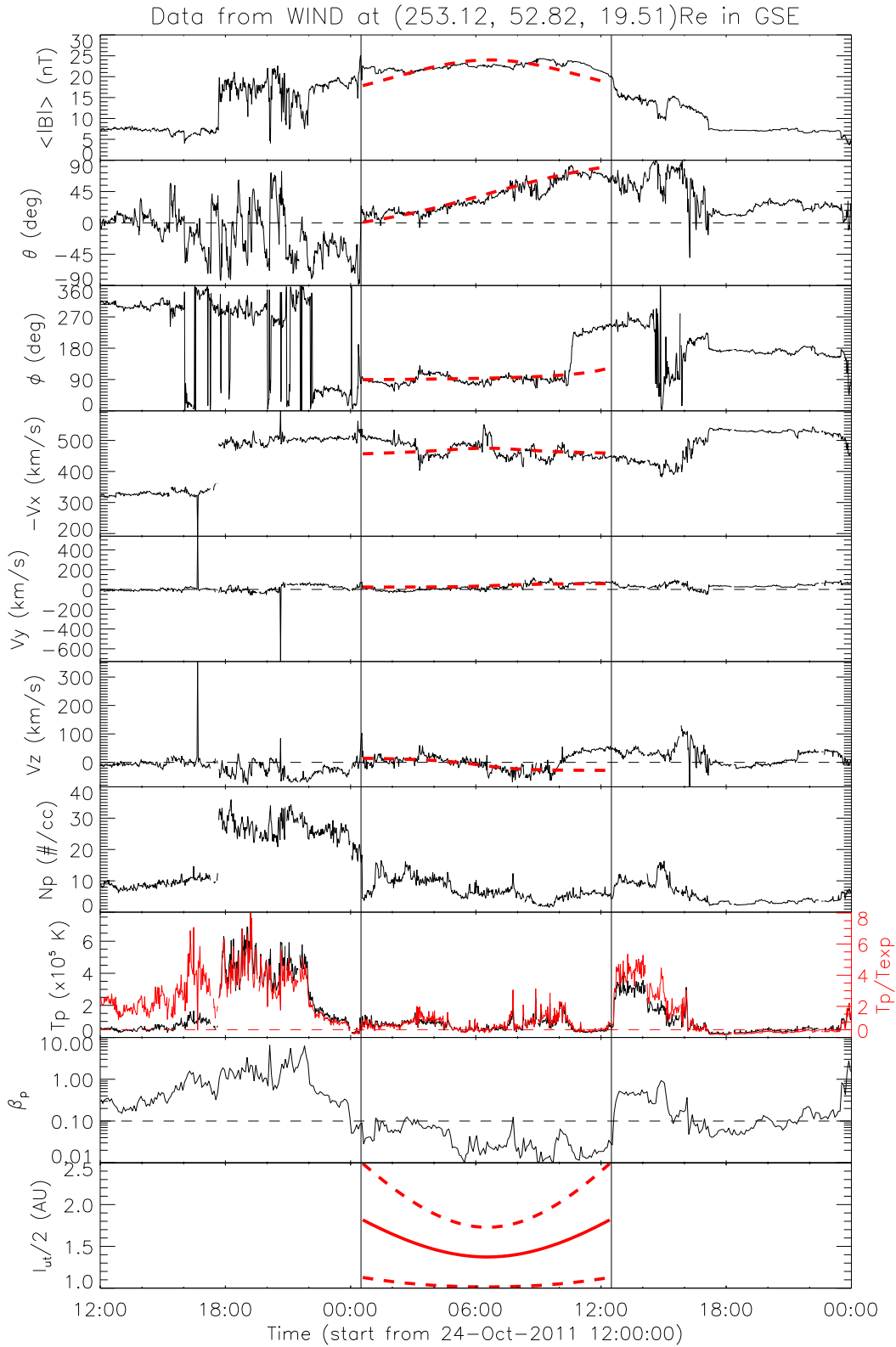}
	\caption{In situ observation of the ICME resulting from the 2011 October 22 solar event. Shown from top to bottom are magnetic field magnitude $B$, field inclination angle $\theta$ (with respect to the ecliptic plane), azimuthal angle $\phi$ (0 deg pointing to the Sun), three components $(V_x,V_y,V_z)$ of the solar wind velocity in the GSE coordinate system, proton density $N_p$, proton temperature $T_p$ (superimposed by $T_p/T_\mathrm{exp}$, where $T_\mathrm{exp}$ is an empirical temperature determined from solar wind speed), plasma $\beta$, and half of the field-line length assuming that the flux-rope axial length ranges between 2--$\pi$ AU. The dashed red curves show the model fitting results with the uniformly twisted flux-rope (Gold-Hoyle) solution. The closest approach of the observational path to the flux-rope axis $d$ is 0.37 in terms of the rope radius; the goodness of fit $\chi_n$ is 0.21, which is the normalized root mean square of the difference between the fitting and observation. \label{fig:mc111022}}
\end{figure*}

\begin{figure*} 
	\centering
	\includegraphics[width=0.8\hsize]{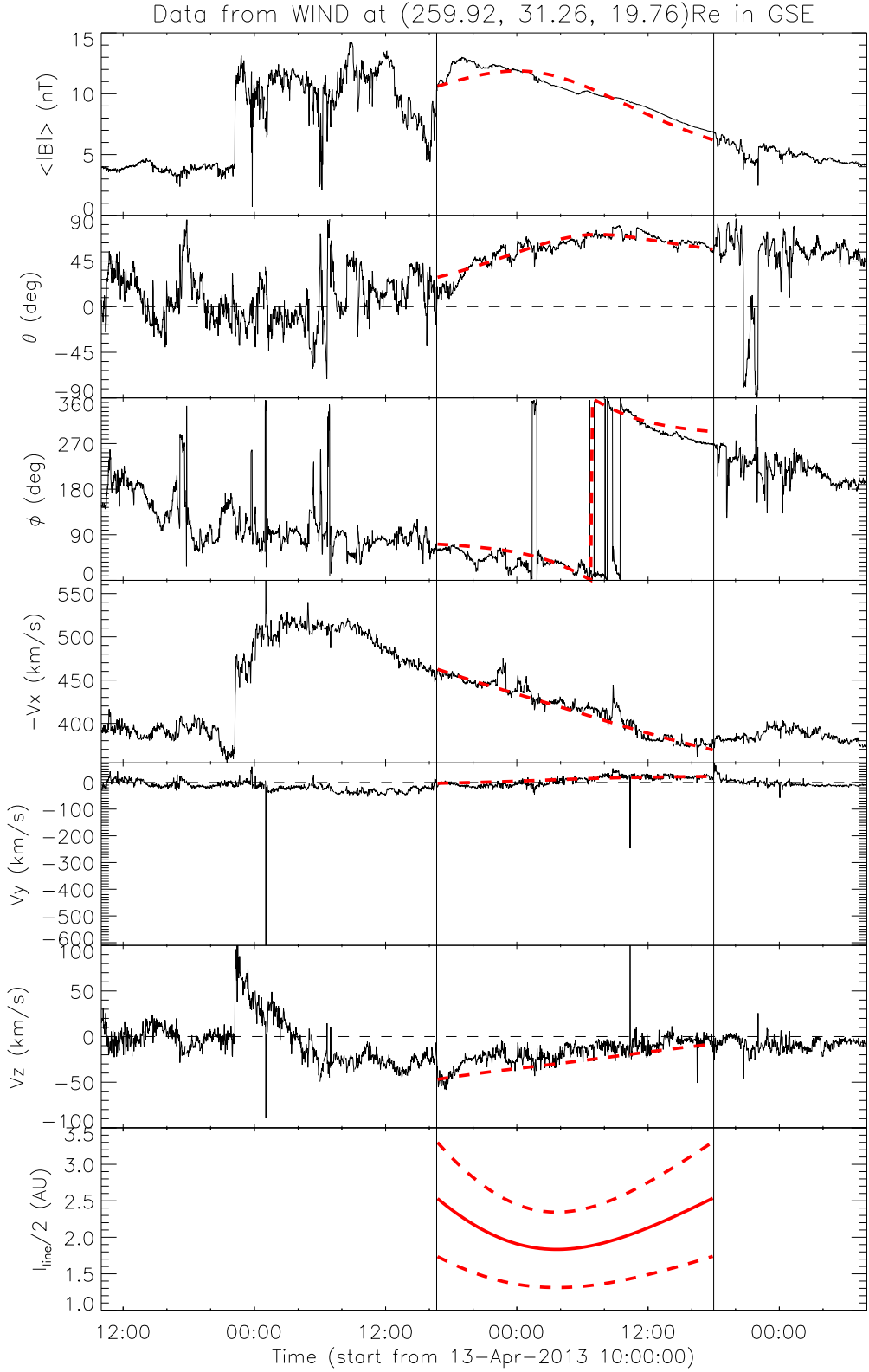}
	\caption{In situ observation of the ICME resulting from the 2013 April 11 solar event. Shown from top to bottom are magnetic field magnitude $B$, field inclination angle $\theta$, azimuthal angle $\phi$ (0 deg pointing to the Sun), and three components $(V_x,V_y,V_z)$ of the solar wind velocity in the GSE coordinate system. The closest approach $d=0.60$, and the goodness of fit $\chi_n=0.18$. \label{fig:mc130411}}
\end{figure*}

\begin{figure*} 
	\centering
	\includegraphics[width=0.8\hsize]{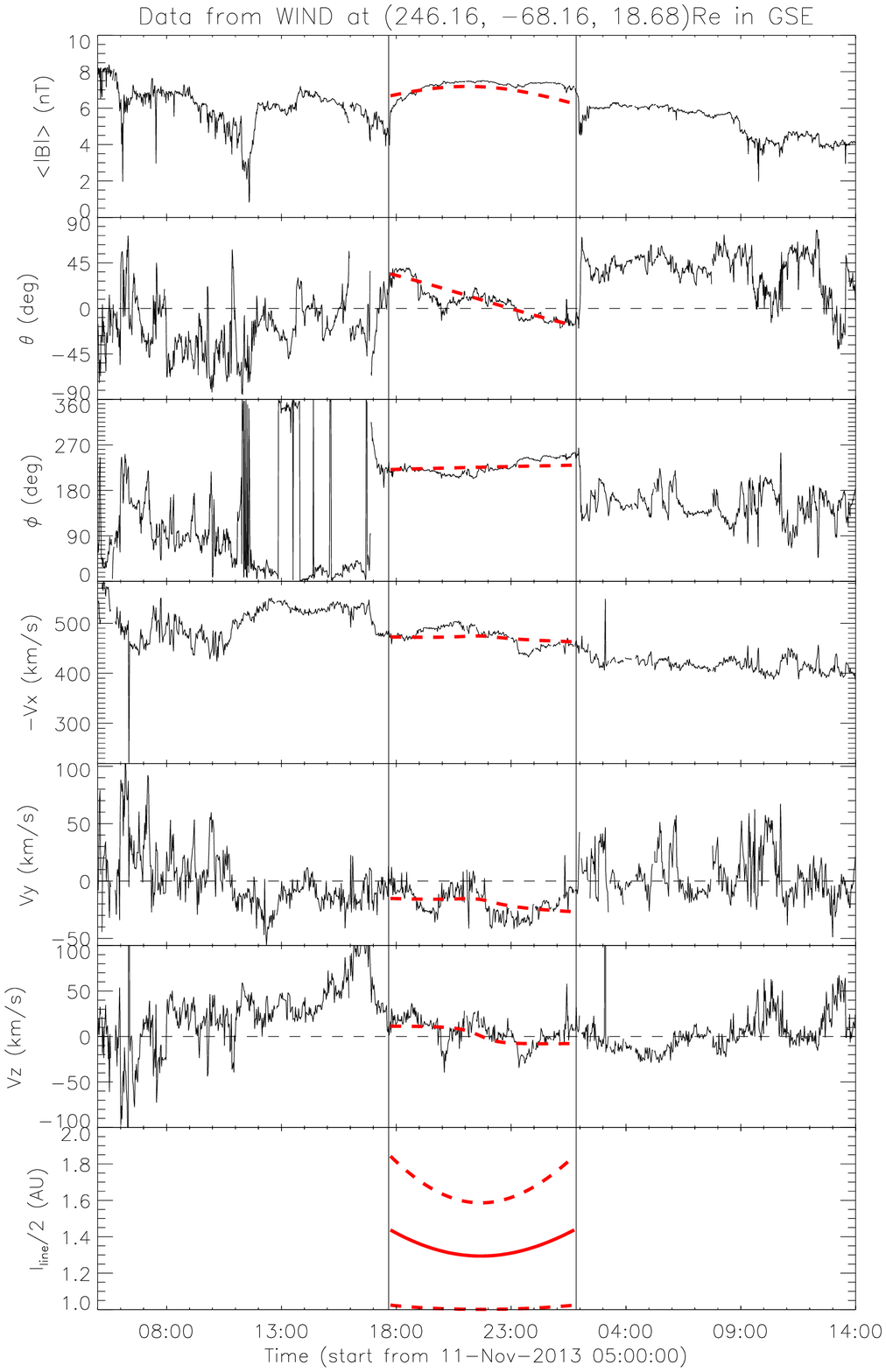}
	\caption{In situ observation of the ICME resulting from the 2013 November 7 solar event. The panels are displayed in the same format as in Figure~\ref{fig:mc130411}. The closest approach $d=0.23$, and the goodness of fit $\chi_n=0.19$. \label{fig:mc131106}}
\end{figure*}

\begin{figure*} 
	\centering
	\includegraphics[width=0.8\hsize]{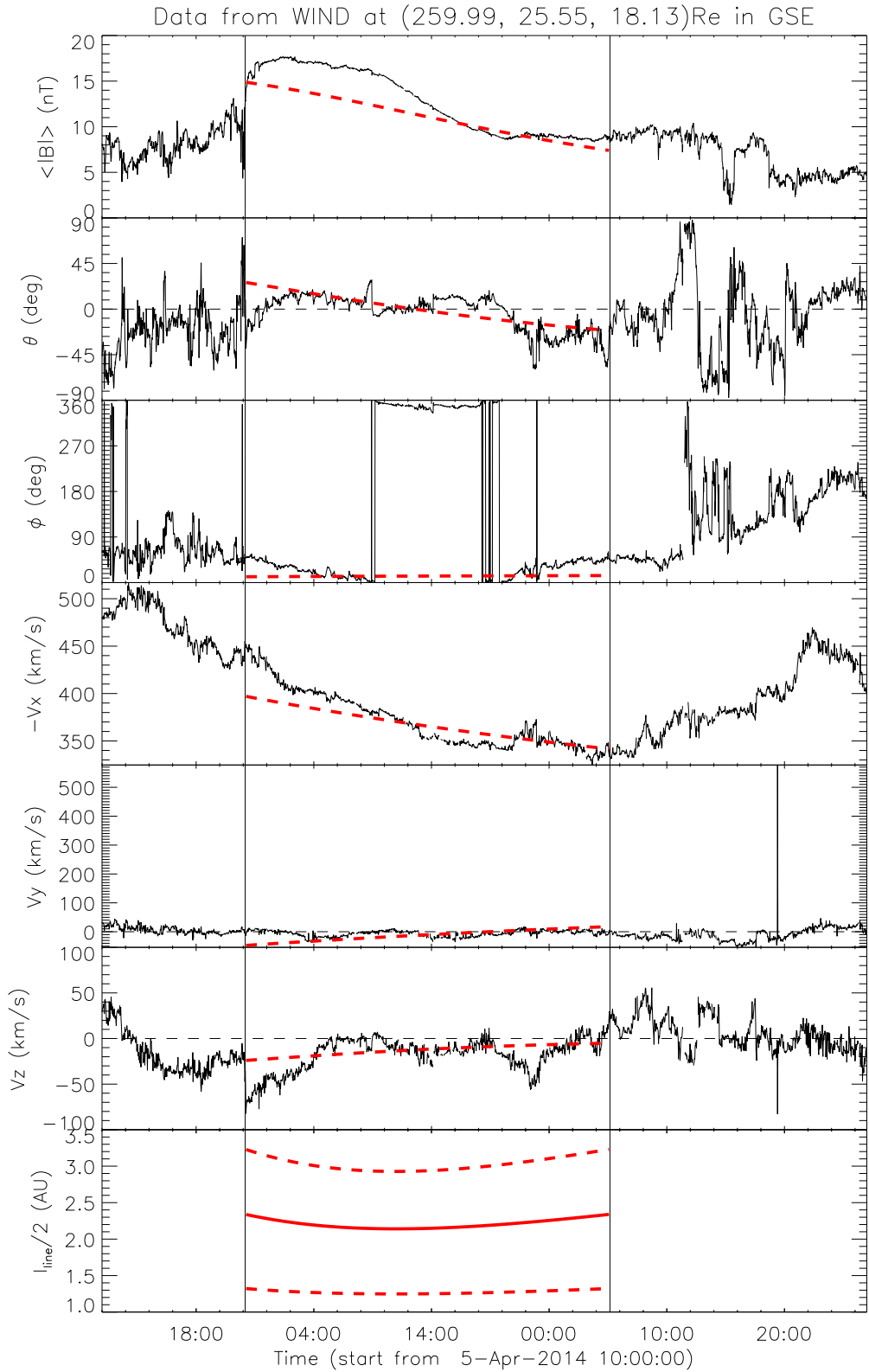}
	\caption{In situ observation of the ICME resulting from the 2014 April 2 solar event. The panels are displayed in the same format as in Figure~\ref{fig:mc130411}. The closest approach $d=0.88$, and the goodness of fit $\chi_n=0.33$.  \label{fig:mc140402}}
\end{figure*}
\end{appendix}

\end{document}